\documentclass{elsart}

\usepackage{epsfig}
\usepackage{amssymb,amsmath}

\begin{document}

\begin{frontmatter}        

\title{Random walk theory of jamming in a cellular automaton model for
traffic flow}

\author[Duisburg]{Robert Barlovi\'{c}},
\author[Cologne]{Andreas Schadschneider} and 
\author[Duisburg]{Michael Schreckenberg}

\address[Duisburg]{Theoretische Physik FB 10, 
	Gerhard-Mercator-Universit\"at Duisburg,
        D-47048 Duisburg, Germany,
        email: barlovic, schreck@uni-duisburg.de}
\address[Cologne]{Institut f\"ur Theoretische Physik, 
	Universit\"at zu K\"oln,
   	D-50937 K\"oln, Germany,
  	email: as@thp.uni-koeln.de}
                                             
\begin{abstract}

The jamming behavior of a single lane traffic model based on a
cellular automaton approach is studied. Our investigations concentrate
on the so-called VDR model which is a simple generalization of the
well-known Nagel-Schreckenberg model. In the VDR model one finds a
separation between a free flow phase and jammed vehicles. This phase
separation allows to use random walk like arguments to predict the
resolving probabilities and lifetimes of jam clusters or
disturbances. These predictions are in good agreement with the results
of computer simulations and even become exact for a special case of
the model. Our findings allow a deeper insight into the dynamics of
wide jams occuring in the model.

\end{abstract}           
\end{frontmatter}


\section{Introduction}    

One of the most interesting phenomena of traffic flow is the
appearance of traffic jams. Two classes of jams can be distinguished:
Those induced by external influences, e.g., bottlenecks, lane
reductions or intersections \cite{daganzo,helbing}, and spontaneous
jams, sometimes called phantom jams, caused by velocity fluctuations.
The latter effect was first shown by Treiterer \cite{treiterer} who
examined a series of aerial photographs of a highway. Further
measurements on real traffic by Kerner and coworkers \cite{kerner2}
have revealed the existence of phase separated jammed states and
homogeneous metastable states with a very high throughput.
Experimentally they observed several characteristic features of wide
jams. Wide jams are compact stable structures which are clearly
separated from moving vehicles. The outflow of a jam is reduced
remarkably compared to the maximal flow in homogeneous traffic. The
upstream velocity of the jam front is approximately constant about
$15~{km}/{h}$, independent of the road conditions. These quantities
are nowadays regarded as important parameters of highway traffic which
can be used for example to calibrate theoretical models. In addition
to homogeneous states and wide traffic jams further phases of
vehicular traffic on highways were found in recent years. Empirical
observations \cite{kerner3,kerner4} showed the existence of so-called
synchronized traffic states and stop and go traffic. Recently Neubert
{\it et al.} \cite{neubert} analyzed single-vehicle data in detail.
They showed that the cross-correlations between density and flow can
be used as a criterion for the identification of the different traffic
states. It should be mentioned that the nature of these traffic states
is still under a vivid debate.

Besides the analysis of experimental data various theoretical
approaches for the description traffic flow theory have been suggested
in recent years \cite{helbing,tgf,tgf97,review}. These approaches can
be divided into fluid-dynamical descriptions and microscopic models
where individual vehicles are represented by a particle. The first
model with the ability to describe the formation and evolution of wide
jams accurately was a macroscopic approach proposed by Kerner and
Konh\"auser \cite{kerner}. Besides these fluid-dynamical models
several microscopic approaches \cite{tgf,tgf97,review} that are able
to reproduce the dynamics of wide jams have been discussed. Nowadays
one of the main interests in applications to real traffic is to
perform real-time simulations of large networks with access to
individual vehicles. Therefore cellular automata (CA) models have
become quite popular for traffic flow simulations.

Due to their simplicity these microscopic models can be used very
efficiently for computer simulations. A few years ago, Nagel and
Schreckenberg (NaSch) proposed a probabilistic CA model for traffic
flow \cite{nagel93}. This model is the simplest known CA that can
reproduce the basic phenomena encountered in real traffic, e.g., the
occurance of phantom traffic jams. On the other hand, the NaSch model
can not explain all experimental results. Therefore modifications have
been suggested to obtain a description on a more detailed level. Here
we concentrate our investigations on the NaSch model with
velocity-dependent randomization (VDR model \cite{robert}). It
exhibits metastable states and large phase separated jams whose
upstream front velocities can be easily adjusted through its
randomization parameters.
 
Recently Knospe {\it et al.} \cite{knospe} have proposed a high
fidelity cellular automaton model for highway traffic which is able to
reproduce empirical single-vehicle data \cite{neubert}. In connection
to the work presented in this paper it is interesting to note that the
formation of wide jams in the model of \cite{knospe} is implemented
using VDR-like mechanisms. An additional interesting application of
the VDR model, or models exhibiting metastability in general, is the
flow optimization by stabilizing homogeneous states of traffic at
bottlenecks \cite{robert,robert2,santen,review}. Due to the wide
field of applications and the fact that the stochastic dynamics of
traffic jams is similar in the different CA approaches there is a need
for an exact understanding of the traffic jams that occur in the
models or, at least, for good approximations.


\section{NaSch model with velocity-dependent randomization VDR}
\label{sec_NaSch}

The description of a physical system in terms of a cellular automata
(CA) approach is in general an extreme simplification of the real
world conditions. In the CA models for traffic the space, speed,
acceleration and even time are treated as discrete variables. The
motion of vehicles is realized through a simple set of rules. In the
spirit of modeling complex phenomena in statistical physics Nagel and
Schreckenberg (NaSch) have chosen a minimal set of rules for their
model \cite{nagel93}. The aim of the NaSch model is to describe basic
phenomena of traffic flow and not to be very accurate on a microscopic
level. Therefore additional rules have to be added or the standard set
of rules has to be modified for a proper modeling of the
fine-structure of traffic flow (see \cite{knospe} and references
therein).

In the NaSch model the road is divided into cells of length $7.5
m$. Each cell can either be empty or occupied by at most one car. The
speed $v_n$ of each vehicle $n=1,2,\ldots,N$ can take one of the
$v_{{\rm max}}+1$ allowed integer values $v_n=0,1,...,v_{{\rm max}}$.
The state of the road at time $t+1$ can be obtained from that at time
$t$ by applying the following rules for all cars at the same time
(parallel dynamics):
\begin{itemize}
\item Step 1: {\it Acceleration:}\\
$v_{n}\rightarrow \min(v_{n}+1,v_{{\rm max}})$
\item Step 2: {\it Braking:}\\
$v_{n}\rightarrow \min(v_{n},d_{n}-1)$
\item Step 3: {\it Randomization with probability $p$:}\\
$v_{n}\rightarrow \max(v_{n}-1,0)$
\item Step 4: {\it Driving:}\\
$x_{n}\rightarrow x_{n}+v_{n}$
\end{itemize}
Here $x_{n}$ denotes the position of the $n$-th car and
$d_{n}=x_{n+1}-x_{n}$ the distance to the next car ahead. The density
of cars is given by $\rho=N/L$, where $L$ is the length of the system,
i.e.\ the number of cells. One time-step corresponds to approximately
$1~s$ in real time.

\begin{figure}[!h]
\centerline{\psfig{figure=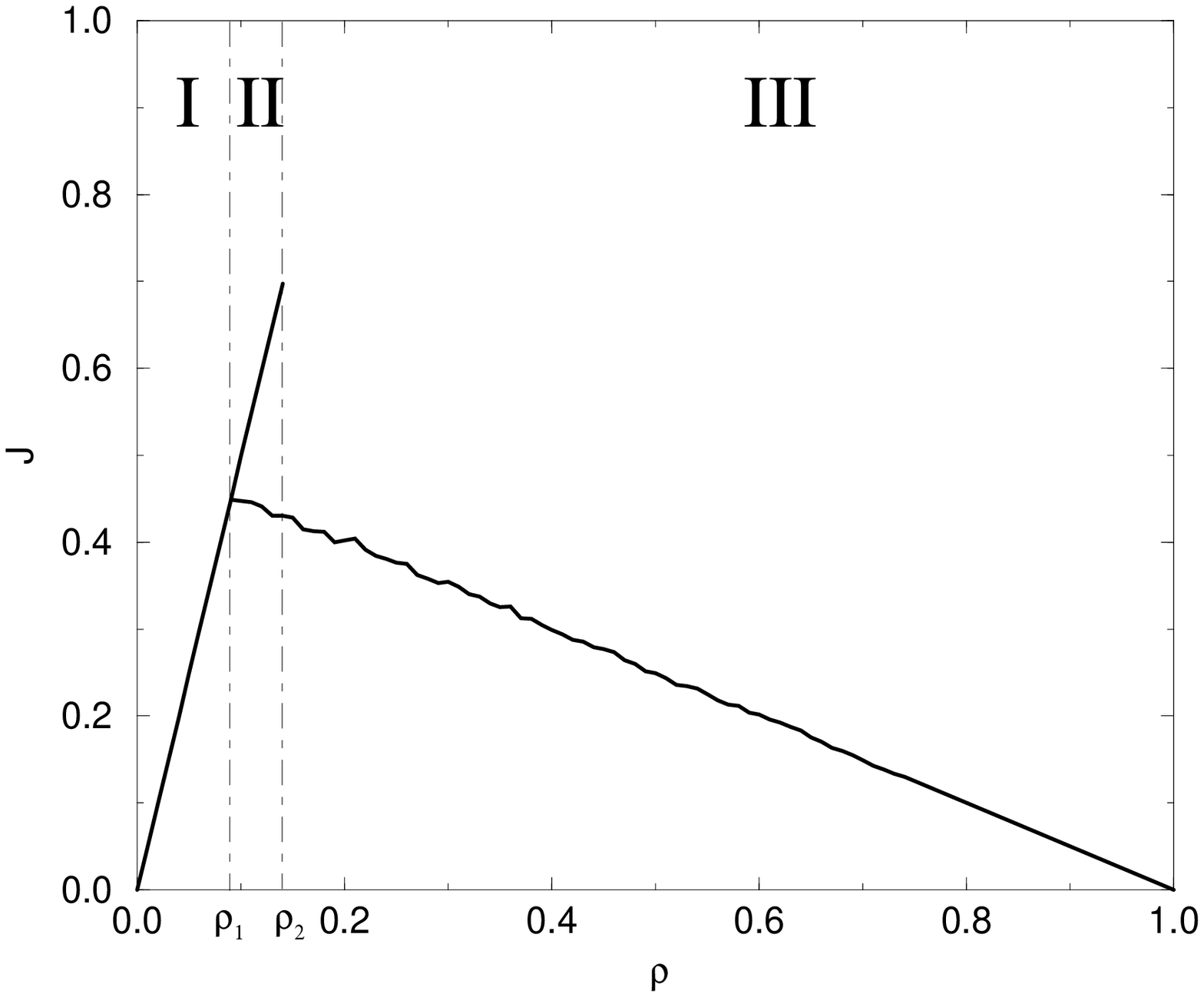,width=6.5cm}\psfig{figure=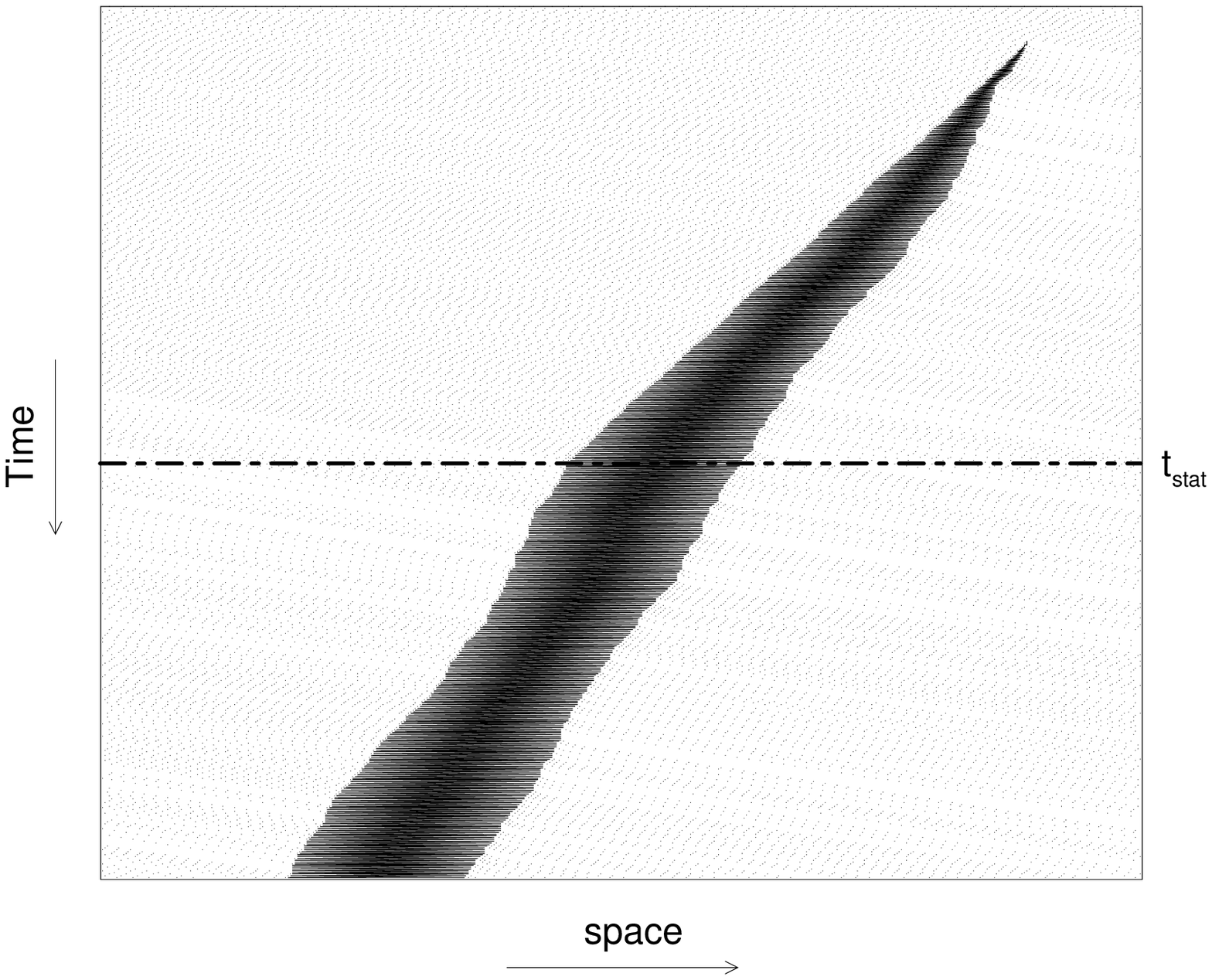,width=6.5cm}}
\caption{\protect{{\bf Left:} Typical fundamental diagram of the VDR
model. The fluctuation parameter of standing cars is set to
$p_{0}=0.5$ and for free flowing vehicles to $p=0.01$. The metastable
branch in regime II can clearly be identified. {\bf Right:} Space-time
diagram of a spontaneous emerging jam in the VDR model for $\rho =
\frac{1}{7}, p_{0}=0.5,p=0.01$. The jam is growing monotonously from
its first appearence until inflow and outflow are equal due to the
periodic boundary condition. After $t_{{\rm stat}}$ timesteps the
average jam length fluctuates around its mean value.}}
\label{fundi}
\end{figure} 

The NaSch model neither exhibits metastable states and hysteresis nor
wide phase-separated jams. However, a simple generalization exists
which is able to reproduce these effects. In the so-called VDR model
\cite{robert} a velocity-dependent randomization (VDR) parameter $p =
p(v(t))$ is introduced in contrast to the constant randomization in
the original formulation. This parameter has to be determined {\em
before} the acceleration step 1. One finds that the dynamics of the
model strongly depends on the randomization. For simplicity we study
the so-called slow-to-start case
\begin{equation}
p(v) = \left \{
        \begin{array}{ccc}
        p_{0} & {\rm \ \ \ \ for} & v = 0, \\
        p & {\rm \ \ \ \ for} & v > 0,
        \end{array}
        \right.           
\label{pvonv}
\end{equation}  
with two stochastic parameters $p_{0}$ and $p$ which already contains
the expected features. In this case the randomization $p_{0}$ for
standing cars is much larger than the randomization $p$ for moving
cars, i.e., $p_{0} \gg p$. An interesting fact is that an alternative
choice of $p(v)$, e.g., $p_{0} \ll p$, leads to a completely different
behavior. Note that for $p_{0} = p$ the original NaSch model is
recovered. Throughout the paper we will always assume that $p\ll
p_{0}$ if not stated otherwise.

Typical fundamental diagrams of the VDR model (see Fig.~\ref{fundi})
show a density regime $\rho_1 < \rho < \rho_2$ where the flow can take
on two different states depending on the initial conditions. The
homogeneous branch is metastable with an extremely long life-time and
the jammed branch shows phase separation between jammed and free
flowing cars. Neglecting interactions among free flowing vehicles the
fundamental diagram of the model can be derived on the basis of
heuristic arguments in good agreement to numerical results
\cite{robert}. The microscopic structure of the jammed states in the
VDR model differs from those found in the NaSch model. While jammed
states in the NaSch model contain clusters with an exponential size
distribution \cite{nageljam,as}, one finds phase separation in the VDR
model. The reason for this different behavior is the reduction of the
outflow of a jam compared to the maximal possible global flow. A large
stable jam can only exist if the outflow from the jam is equal to the
inflow. If the outflow is the maximal possible flow a stable jam can
only exist at the corresponding density $\rho_{{\rm max}}$, but will
easily dissolve due to fluctuations. If the outflow of a jam is
reduced, the density in the free flow regime is smaller than the
density $\rho_{{\rm max}}$ of maximal flow where interactions between
vehicles lead to jams so that cars can propagate freely (for $p\ll
p_0$) through the low density part of the road. Therefore no
spontaneous jam formation is observable. It is obvious that no phase
separation can occur in the NaSch model due to the fact, that the
outflow of a jam is maximal. Figure~\ref{fundi} shows the typical
structure of the jammed state in the VDR model.


\section{Jamming process}
     
The fundamental diagram of the VDR model with periodic boundary
conditions can be divided into three different regimes according to
the jamming properties. For densities up to $\rho_{1}$ no jams with
long lifetime appear and jams existing in the initial conditions
dissolve very quickly. Obviously in the regime $0\leq \rho\leq
\rho_1$ the outflow of a jam is greater than the inflow. This
behavior has to be contrasted to the one found for densities above
$\rho_{2}$. Here no homogeneous state without jams can exist. The
most interesting regime lies between the two densities $\rho_{1}$ and
$\rho_{2}$ where the system can be in two different states. One is a
metastable homogeneous state with an extremely long life time where
jams can appear due to internal fluctuations. Note that the
homogeneous states can be destroyed by external perturbations, e.g.,
by stopping cars. The other state is a phase separated state with
large jams which can be reached through the decay of the homogeneous
state or directly owing to the initial condition. The microscopic
structure of a spontaneous emerging jam can be seen in the space time
plot in Fig.~\ref{fundi}. The origin of the wide jam in the initially
homogeneous state is a local velocity fluctuation that leads to a
stopped car.
A velocity fluctuation of a car can force the following car to brake
if its headway becomes too small. If the local density is large enough
this can lead to a chain reaction which finally forces a car to stop.
Such local velocity fluctuations determine the typical density-dependent 
lifetime of the homogeneous states.

As one can see in Fig.~\ref{fundi}, the density in the outflow regime
of the jam is reduced compared to the average density. Therefore the
jam length is growing approximately linear until outflow and inflow
coincide due to the periodic boundary conditions. It is quite evident
that interactions between vehicles in the outflow region of a jam are
negligible so that no spontaneous jams can appear. A recent study by
Pottmeier {\it et al.} \cite{potti,potti2} indicates that local
defects can break the strong phase separation found in the VDR model
in such a way that stop and go traffic can exist.

Furthermore Fig.~\ref{fundi} shows that strong fluctuations in the
upstream jam front appear due to the fact that the outflow of a jam is
determined by a stochastic parameter. As aforementioned the locally
emerged jam will probably grow for a certain time because the mean
inflow into the jam is larger than the mean outflow. At this point it
should be stressed, however, that although the inflow is larger than
the outflow these quantities are stochastic and therefore even in this
growth regime a complete dissolution of the emerged jam is possible
through fluctuations. The dynamics of this growth (dissolution)
process is the main object of our investigations in this paper.
Nonetheless, assuming that the locally emerged jam does not dissolve
it will grow a certain time $t_{\rm stat}$ until the mean inflow and
outflow are equal due to periodic boundary conditions. The average jam
length then strongly fluctuates. This can also lead to a complete
dissolution of the jam when these fluctuations are of the order of the
jam length.  The phase separation in the jammed state can be
identified directly using the jam-gap distribution
\cite{georg,georg2}. It was shown that for $p=0$ the jam is compact in
the sense that no holes between the jammed cars appear. For $p>0$
holes are formed due to velocity fluctuations of vehicles entering the
jam. Nevertheless, it can be assumed that for $p_{0}\gg p$ the jammed
states are phase separated, i.e. the size of the jam is of the order
of the system size.

\begin{figure}[h]
 \centerline{\psfig{figure=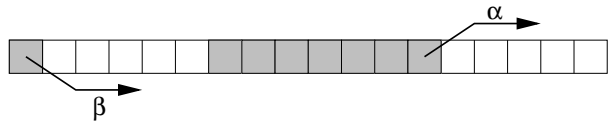,width=8cm}}
 \caption{\protect{Schematic representation of a single jam. Cars are
represented by grey cells and empty cells are white. $\alpha$ is the outflow
of the jam. The inflow $\beta$ is generated by the outflow of a megajam
at the left end.}}
\label{jam}
\end{figure}

A schematic representation of a single jam is depicted in
Fig.~\ref{jam} where jammed cars are represented by grey cells and
empty cells are white. The parameter $\alpha$ denotes the outflow and
is equal to $p_{0}$ in the studied model. Since for $p\ll p_0$ car-car
interactions can be neglected in the outflow region a car that leaves
a jam can be treated as escaped for all times. Therefore the velocity
of the upstream jam front is only determined by $p_{0}$, i.e., the
waiting time of the first car. Neubert {\it et al.}\ \cite{neubert2}
analyzed the jam velocity in the VDR model with an autocorrelation
function and find an excellent agreement with the simple assumption
made above. Note that this behavior is quite different from that of
the NaSch model where even in the outflow region spontaneous jams
occur for sufficiently large $p$ (stop and go traffic). The parameter
$\beta$ is a measure for the inflow into the jam. This quantity is
determined by the vehicle distribution on the road section upstream
the jam (see Fig.~\ref{jam}).

In the next section we will introduce a stochastic theory of jamming
based on random walk like arguments. Our approach gives the exact
solution in the special case $p=0$ where the fluctuations of free
flowing vehicles are completely suppressed. For the general case
$0<p\ll p_0$ the results are in good agreement with simulations.


\section{Random walk theory of jamming}
\label{sec_RW}
           
The main object of our investigation is a sequence of $n$ cars at rest
forming a compact jam (see Fig.~\ref{jam}). In every timestep the
first car can leave the megajam with probability $\alpha = p_{0}$
according to the acceleration and randomization steps in the update
algorithm (see Sec.~\ref{sec_NaSch}). Additionally a new car is able
to enter the jam at its end with probability $\beta$ which is
determined by the vehicle distribution behind the jam. In general
$\beta$ will be generated by the outflow of a megajam (see below).

The outflow $\alpha$ realized through $p_{0}$ is an independent
identically distributed (i.i.d.) random variable. For simplicity we
start with the case that $\beta$ is also an i.i.d.\ random variable.
This condition is fulfilled if fluctuations of free flowing vehicles
are completely suppressed, i.e., for $p=0$, and when the vehicle
distribution behind the considered jam is generated by the outflow of
a megajam. In this case we will consider open boundary conditions with
an (infinite) megajam at the left end and an empty system at the right
end (Fig.~\ref{jam}). The density in the free flow regime is then
determined by the waiting time of the first car in the megajam. In
this special case the gaps between the vehicles are always a multiple
of their maximum velocity $v_{{\rm max}}$ since the waiting times are
discrete and no interactions between cars occur in the outflow region
for $p=0$. For example a waiting time of $t_w$ timesteps leads to a
gap of $t_w\cdot v_{{\rm max}}$ cells. Then a car with this gap will
reach the end of the jam exactly $t_w$ timesteps after its
predecessor. It is obvious that the waiting times of the first car in
the megajam determine directly the inflow $\beta$ into the analyzed
jam which therefore can be treated as an i.i.d.\ variable.  Note that
the randomization parameter $\tilde{p}_0$ of cars standing in the
megajam can be considered as a control parameter for the vehicle
distribution in the free flow regime, i.e., the inflow
$\beta=\beta(\tilde{p}_0)$.  Therefore it is in general different from
the $p_{0}$ of the cars that have already left the megajam, i.e., the
$p_0$ governing the resolution of the jam under consideration. In
this way we can realize $\alpha \ne \beta$ for the open system. The
case of periodic boundary conditions can be treated as a special case
of the problem where in the jammed state the inflow into the jam is
determined by the outflow out of the jam and hence $\alpha=\beta$. A
detailed description of the initial and boundary conditions considered
here as well as a discussion of the influence of deviating gap
distributions in the inflow region (e.g.\ if $\beta$ is no longer an
i.i.d.\ variable) will be given below.

In the following we will map the jam dynamics onto a random walk
problem. The number of standing cars $n$ in the jam determines the
position $n$ of a random walker. The walker moves on a discrete
lattice in discrete time. A car leaving (entering) the jam then
corresponds to one step to the left (right). In the following we
determine the probability $\pi_{t,n}$ that a jam of width $n$ resolves
after $t$ timesteps. Here a jam is considered to have resolved when
the last remaining car accelerates.

In random walk terminology this problem is equivalent to the
calculation of the first passage time of a walker starting at position
$n_0=n(t=0)$. $\pi_{t,n}$ is the probability that walker at position
$n$ reaches the origin $n=0$ of the system in $t$ timesteps.  Taking
into account that $\alpha$ and $\beta$ are i.i.d.\ random variables
one gets the following master equation for the duration of the process
until the random walker reaches the origin:
\begin{eqnarray}
\pi_{t+1,n>1} & = & \alpha (1-\beta) \pi_{t,n-1} + \beta (1-\alpha) 
  \pi_{t,n+1} + \left[(1-\alpha)(1-\beta) + \alpha\beta\right]
  \pi_{t,n},\nonumber\\  
 \pi_{t+1,1} & = & \alpha \pi_{t,0} + \beta (1-\alpha) \pi_{t,2} +
 (1-\alpha)(1-\beta) \pi_{t,1}.
\label{system}
\end{eqnarray} 
It is obvious that $\pi_{0,n_0>0} = 0$ and that by definition
$\pi_{1,1} = \alpha$ since a jam of length $n=1$ resolves with
probability $\alpha$ in one timestep independent of the
inflow\footnote{This is due to the update procedure and will be
discussed later.}. Furthermore we assume an absorbing barrier at the
origin $\pi_{0,0} = 1$, i.e.\ the process stops when the random walker
reaches the origin (the jam is dissolved). Note, that the variable $n$
covers the whole spektrum of possible positions of a random walker
during his movement, while $n_0$ denotes the starting position. For a
discussion of the mathematical aspects of first passage time problems
we refer to \cite{vanKampen,Feller}.

As an example for the derivation of these equations consider a walker
starting at position $n_0>1$. If the first trial results in a movement
to the left the process continues as if the initial position had been
$n_0-1$. A movement to the left means that a jam of width $n_0$
evolves into a jam of width $n_0-1$. This event occurs if the first
car leaves the jam and no additional car enters it. The probability
for this is $\alpha (1-\beta)$. Similarly, if the first trial results
in a movement to the right the process continues as if the initial
position had been $n_0+1$. The probability for a movement to the right
is given by $\beta (1-\alpha)$, i.e., the first car remains in the jam
and an additional car enters it. Furthermore the position of the
walker, and thus the jam length, are unchanged if either one car
leaves and one car enters the jam (probability $\alpha\beta $) or no
car leaves and enters the jam (probability $(1-\alpha)(1-\beta)$).
From the first equation of (\ref{system}) we see that the random walk
is symmetric (for $n>1$) if $\alpha(1-\beta)=(1-\alpha)\beta$. For
$\alpha(1-\beta)> (1-\alpha)\beta$ the walker is biased to the left
and the jam will dissolve quickly. For $\alpha(1-\beta)<
(1-\alpha)\beta$ the bias is to the right and the jam will grow on
average. But even in this case a complete resolution is possible
through fluctuations.

In order to determine the probabilities we introduce the following
generating functions:
\begin{equation}
\Pi_{n_{0}}(z) = \sum_{t=0}^{\infty} \pi_{t,n_{0}}z^{t}.   
\label{generatingone}
\end{equation}  

At this point it has to be taken into account that the case $n=1$ must
be viewed separately.  For a jam consisting of only one standing car
the probability for resolving is equal to $\alpha$ (the standing car
accelerates) even if a new car enters the minijam at its end.
Therefore we will first look at the special case of a random walker
starting at position $n_0=1$.  A typical process in which the jam
resolves after $t>1$ timesteps may be described as follows.\\ (I) The
walker does not move for $\mu$ timesteps.\\ (II) Then the random
walker moves one unit to the right.\\ (III) After the step to the
right the walker first has to return to $n=1$ before the jam resolves,
i.e.\ before the walker reaches $n=0$.  The return to the initial
position $n=1$ will take $\nu - 1$ further timesteps ($\nu =
2,3,\ldots$).\\ (IV) Now there are $t - \nu - \mu$ further timesteps
left for the walker to reach the origin at last.\\ These three events
are mutually independent and so the probability of the simultaneous
realization of the three events is given by the product of the single
probabilities.

We first look at the case $\mu=0$. The cases $\mu>0$ will later be
taken into account iteratively. We start with the solution of event
(III). Since in this case the position of the walker is always larger
than one only the first equation of the system~(\ref{system}) has to
be considered. In contrast to the general solution for event (III) the
hopping probabilities do not depend on the position anymore. A random
walker has to return to the origin (in this case $n=1$) starting from
a position shifted one step to the right (here $n=2$). The solution
of this homogeneous first passage time problem is described in
\cite{vanKampen,Feller}. Assuming an absorbing barrier and
introducing a separate generating function for event (III),
\begin{equation}
\tilde{\Pi}(z) = \sum_{t=0}^{\infty} \tilde{\pi}_{t,1} z^{t},   
\label{generatingtwo}
\end{equation}  
where $\tilde{\pi}_{t,1}$ is the probability that a walker starting at
$n=2$ reaches $n=1$ for the first time after $t$ timesteps, the
following solution for part (III) of the problem can be obtained:
\begin{eqnarray}
\tilde{\Pi}(z) & = &
\frac{1-\left[\alpha\beta+(1-\alpha)(1-\beta)\right]z}{2\beta
(1-\alpha)z} \nonumber \\
& & -
\sqrt{\left[\frac{1-\left[\alpha\beta+(1-\alpha)(1-\beta)\right]z}
{2\beta(1-\alpha)z}\right]^{2} - \frac{\alpha(1-\beta)}{\beta(1-\alpha)}}.    
\label{solutionforgeneratingtwo}
\end{eqnarray}  
The return probability, i.e.\ the probability that the walker reaches
$n=1$ at an arbitrary time, is then given by
\begin{equation}
\tilde{\Pi}(1) = 
        \begin{cases}
        \frac{\alpha(1-\beta)}{\beta(1-\alpha)} & {\rm if\ } 
          \alpha(1-\beta) \leq \beta(1-\alpha) \\
        1 & {\rm if\ } \alpha(1-\beta) \geq \beta(1-\alpha).
        \end{cases}
\label{generatingzone}
\end{equation}  
As already discussed above, for $\alpha(1-\beta)> \beta(1-\alpha)$ the 
walker is biased to the left and will always reach $n=1$. For
$\alpha(1-\beta)< \beta(1-\alpha)$, on the other hand, the walker is
biased to the right and will only return with probability
$\frac{\alpha(1-\beta)}{\beta(1-\alpha)}<1$. 

Now the complete event (I)--(IV) occurs for some $\nu < t$. Summing
over all possible $\nu$ one gets:
\begin{eqnarray}
\pi_{t,1} & = &\beta (1-\alpha) \left[ \tilde{\pi}_{1,1}\pi_{t-2,1} +
 \tilde{\pi}_{2,1}\pi_{t-3,1} + ..... +  \tilde{\pi}_{t-2,1}\pi_{1,1}
 \right]\nonumber\\
 && +  (1-\alpha)(1-\beta) \pi_{t-1,1}.
\label{solutionforgeneratingoneone}
\end{eqnarray}     
The last term takes into account event (I) for $\mu\neq 0$ since
$(1-\alpha)(1-\beta)$ is the probability that a walker at site $n$
will not move. In the first term, $\beta (1-\alpha)$ is the
probability of event (II). The quantity within the brackets is the
probability of events (III) and (IV) for the allowed values of
$\nu$. Note that it is the $(t-1)$-st term of the convolution
$\{\pi_{t,1}\}\ast\{\tilde{\pi}_{t,1}\}$ (see \cite{Feller}). After
multiplying (\ref{solutionforgeneratingoneone}) with $z^{t}$ and
summing over all times one finds an expression for the generating
function (\ref{generatingone}) for $n_0=1$:
\begin{equation}
\Pi_{1}(z) = \frac{\alpha z}{1-(1-\alpha)(1-\beta)z-\beta(1-\alpha)
\tilde{\Pi}(z)z}.   
\label{solutionforgeneratingonetwo}
\end{equation}  
The probability for the complete return 
is given by $\Pi_{1}(1)$. Using equation~(\ref{generatingzone})
it is easy to obtain this quantity explicitly. The solution for the
nontrivial case $\alpha(1-\beta) < \beta(1-\alpha)$ is given by 
\begin{equation}
\Pi_{1}(1) = \frac{\alpha}{\beta}.
\label{generatingztwo}
\end{equation}

We now turn to the general case where one has to deal with starting
positions greater than one, i.e.\ initial conditions consisting of
more than only one standing car, $n_0>1$. Similarly to the foregoing
approach the process of reaching the origin can be seen as the
realization of mutually independent events. For instance, the
probability that a random walker starting at position $n_0=2$ reaches
the origin is given by the event that the walker reaches first
position $n=1$ and thereafter reaches position $n=0$.

\begin{figure}[h]
 \centerline{\psfig{figure=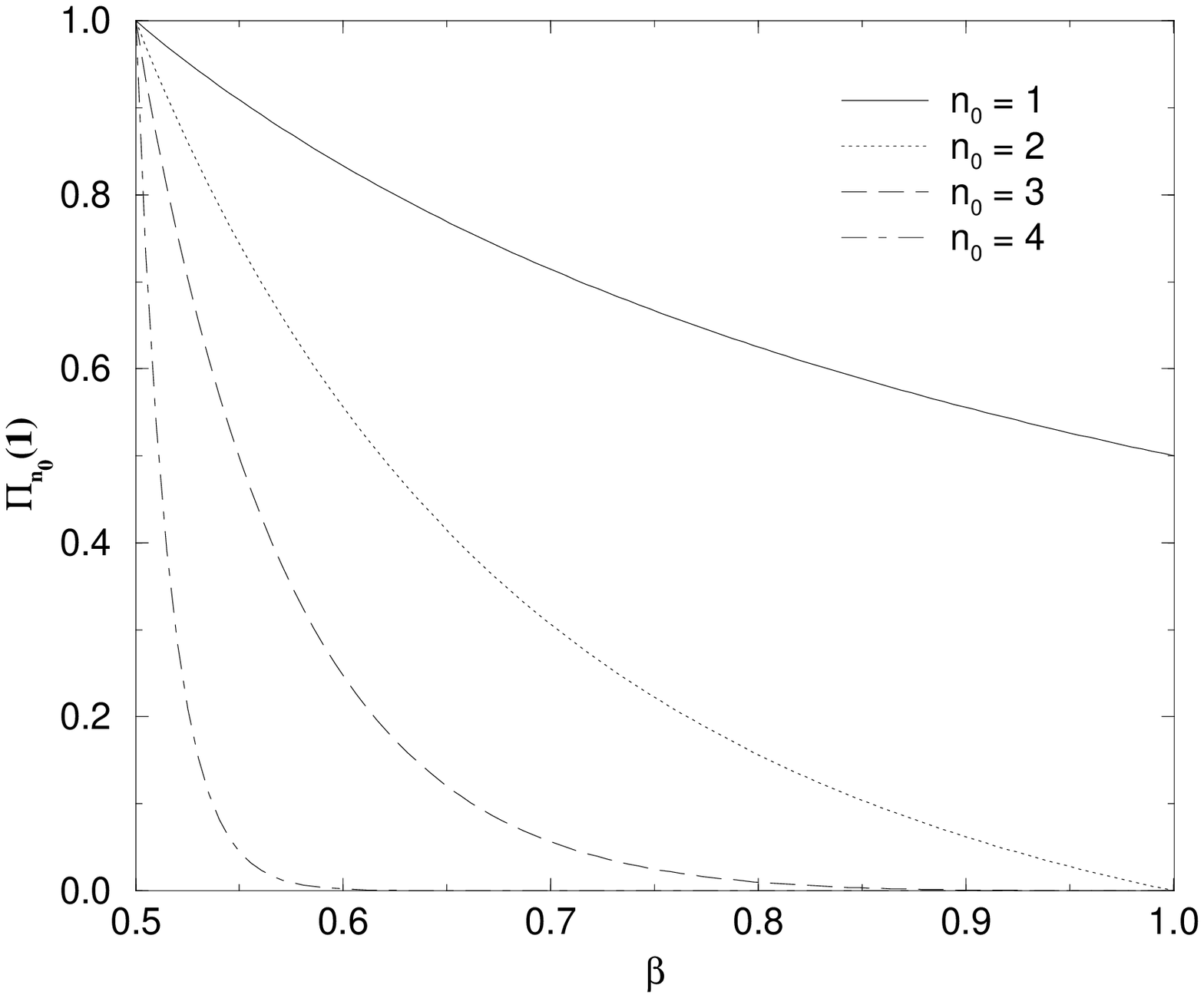,width=6cm}
\psfig{figure=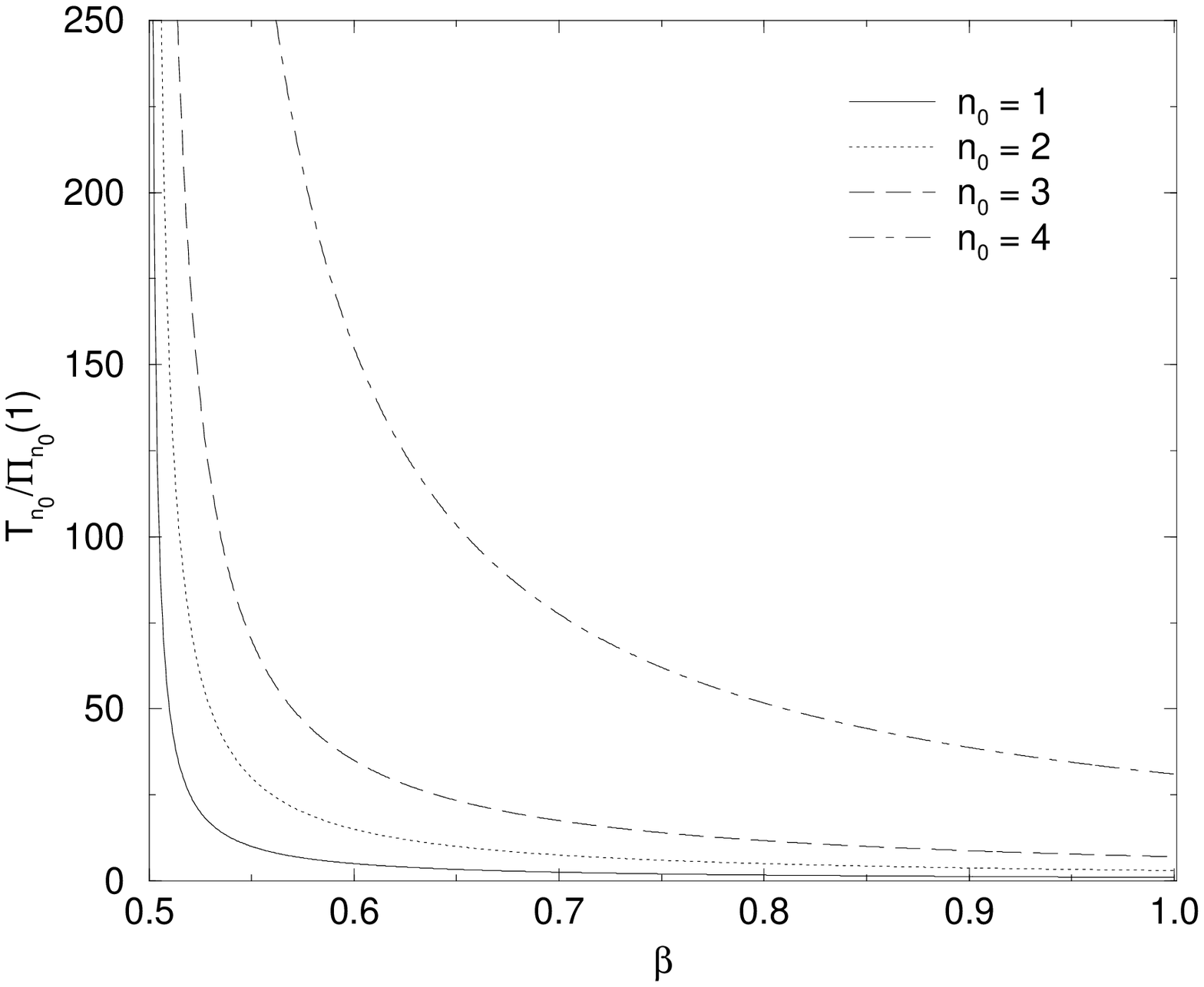,width=6cm} } \caption{\protect{{\bf Left:} The
dissolving probability of a jam (see Eqn.~(\ref{endsolution})) is
illustrated for different $n_0$. It can clearly be seen that the
dissolving probability strongly depends on the initial width $n_0$ of
the jam, and that for the special case $n_0=1$ this quantity will
never fall below $\alpha$.  {\bf Right:} The conditional mean
dissolving time $T_{n_0}/\Pi_{n_0}(1)$ is shown for the set of $n_0$
used in the left part of the diagram. Note that the higher the inflow
$\beta$ the lower the mean dissolving time, but the probability for
dissolution on short times will shrink drastically. The outflow rate
in both diagrams is set to $\alpha = 0.5$. }}
\label{jamandtime}
\end{figure}

The general resolving process of a jam with an initial width $n_0$
standing cars can be described as a chain of processes leading finally
to the case $n=1$. Thus using the generating
functions~(\ref{generatingone}) and~(\ref{generatingtwo}) this
convolution of events can be expressed through
\begin{equation}
\Pi_{n_0}(z) = \tilde{\Pi}(z)^{n_0-1}\Pi_{1}(z).
\label{mostgeneralgenerating}
\end{equation}
With equation~(\ref{generatingzone}) and~(\ref{generatingztwo})
the following relation for the resolving probability of a jam of 
width $n$ can be obtained:
\begin{equation}
\Pi_{n_0}(1) =
\frac{\alpha}{\beta}
\left[\frac{\alpha(1-\beta)}{\beta(1-\alpha)}\right]^{n_0-1}.
\label{endsolution}
\end{equation}
Besides the resolving probability there is also another interesting
quantity that is directly accessible through the generating
function~(\ref{mostgeneralgenerating}), namely the average lifetime
$T_{n_0}$ of a jam of initial length $n_0$:
\begin{equation}
T_{n_0} = \sum_{t=0}^{\infty} t \pi_{t,n_0} = \Pi_{n_0}'(1).   
\label{meantime}
\end{equation}
Using (\ref{mostgeneralgenerating}) one can obtain an explicit result
for $T_{n_0}$. Figure~\ref{jamandtime} shows results for the resolving
probability (\ref{endsolution}) and for the lifetime (\ref{meantime})
of a jam for various initial widths $n_0$. The outflow parameter
(fluctuation parameter of standing cars) is chosen such that
$\alpha=0.5$ in both diagrams. In the left figure a strong dependence
on the starting position of the walker (initial width of the jam) can
be seen. Furthermore one can observe directly an outstanding
difference between the case $n_0=1$ and $n_0>1$. While for $n_0>1$
the resolving probability fastly converges to zero for increasing
$\beta$, this value is shifted to $\alpha$ for $n_0=1$. This shift can
be explained through the update procedure of the model discussed
above. A jam consisting of only one standing car resolves with
probability $\alpha$ even if a new car enters the minijam at its
end. The right part of Fig.~\ref{jamandtime} shows the mean disolving
time for different $n_0$. It is obvious, that this quantity grows with
increasing $n_0$ due to the fact, that the resolving process of a jam
with $n_0$ standing can be described as a chain of resolving processes
of smaller jams. Additionally it should be noted that a higher inflow
$\beta$ leads to lower dissolving times, but it must be taken into
account that the dissolution of a jam under a high inflow $\beta$ is a
rather rare event.

\section{Numerical results}

In the previous section we have derived an analytical expression for
the dissolution probability $\Pi_{n_0}(1)$ of a jam with an initial
width of $n_0$ standing cars. To compare the analytical predictions
with simulation results we use a damage scenario by initializing a
finite jam into an undisturbed system. The reaction of the system to
such disturbances is characterized by the sensitivity $S = 1 -
\Pi_{n_0}(1)$. The sensitivity is simply the probability that a
cluster of $n_0$ standing cars causes a wide jam (for open boundary
conditions) or leads to the jammed state of the system (for periodic
boundary conditions). The following initial conditions are considered
to produce an area of free flowing vehicles in the density regime
$\rho_{1}<\rho<\rho_{2}$. Remember that the mean inflow into the
induced minijam will always be greater than the mean outflow in this
area.

(A) Here we set the randomization parameter $p$ of moving cars to
zero, i.e.\ fluctuations in free flow are suppressed. We choose an
open system with a sufficiently large megajam at the left boundary in
this scenario. The gap distribution is than realized through the
outflow of this megajam and fluctuations of free flowing vehicles are
completely suppressed. The randomization parameter $\tilde{p}_0$ of
standing cars in the megajam is used as control parameter for
$\beta$. After a car has left the megajam it is therefore reset to
$p_0$. Thus the distance between two consecutive cars is always a
multiple of $v_{{\rm max}}$ whereby the inflow into the damage is an
i.i.d.\ random variable in that case in correspondence to the
theory.\\ (B) Also in the second case the free-flow car distribution
is generated through a megajam, but fluctuations are permitted, i.e.,
$p>0$. Hence the gap distribution contains values deviating from
$n\cdot v_{{\rm max}}$ due to velocity fluctuations or braking
events.\\ (C) Finally, a homogeneous initialization of cars in a
periodic system is considered. Simulations for this situation are
performed until the system has relaxed into its free-flow steady
state. As a result of the dynamics this steady state shows large
deviations in the gap distribution in comparison to the megajam
initializations.\\ Note that the inflow into the induced minijam is
controlled in the cases (A)+(B) through the randomization parameter of
standing cars at the left boundary (megajam) while in case (C) the
inflow is controlled indirectly through the density of the homogeneous
initialization. The inflow $\beta$ can be obtained easily in the
simulations.

After the free-flowing vehicles are initialized according to the
scenarios described above we induce a damage in the system by setting
the velocity of a randomly chosen vehicle to zero. The acceleration of
this stopped car is suppressed in the following update steps until the
damage grows to $n_0$ cars. Now, the system is updated according to
the update rules without any further outer influences. The damage
considered can either grow to a large jam or dissolve.

\begin{figure}[h]
 \centerline{\psfig{figure=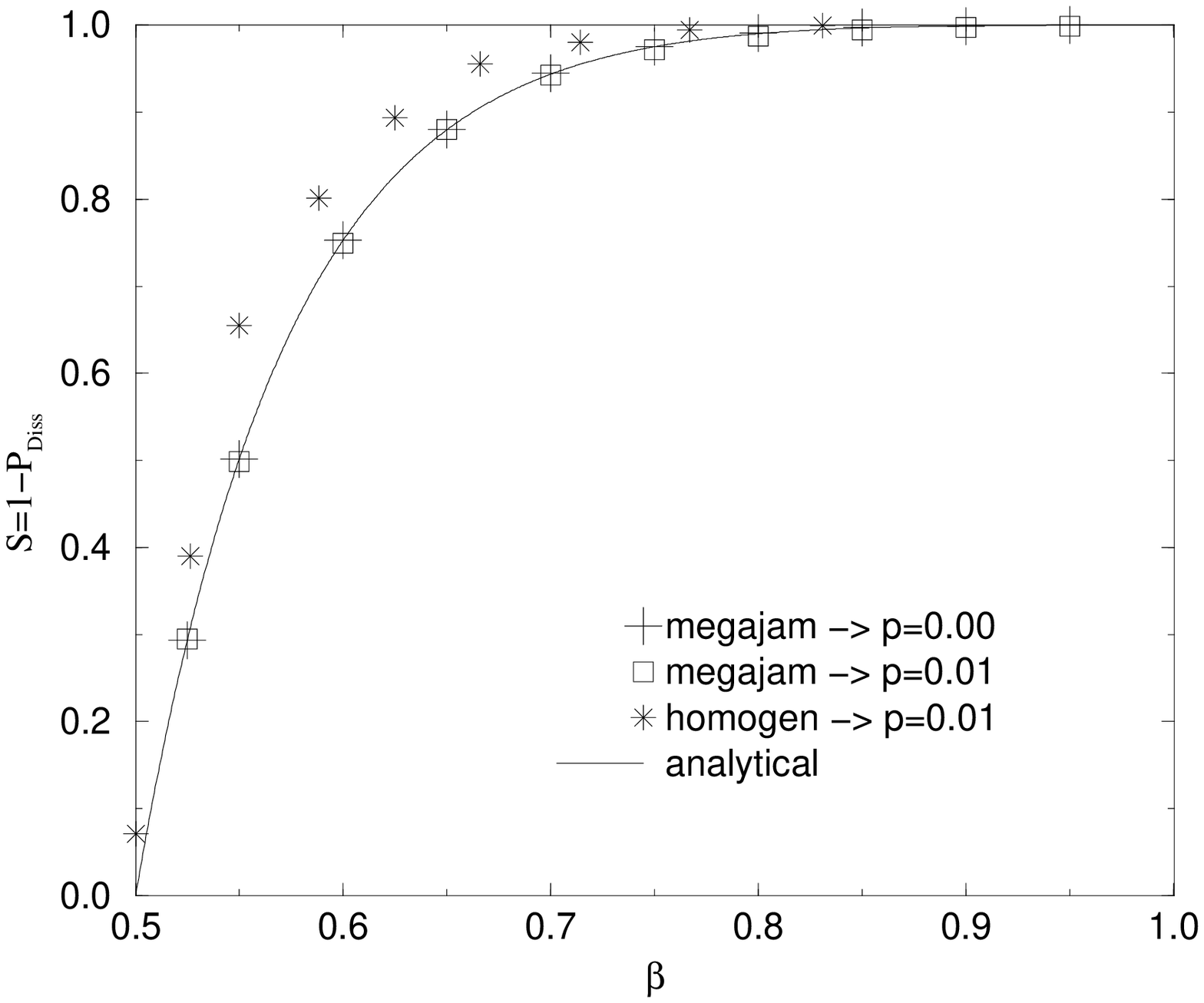,width=6.5cm}
\psfig{figure=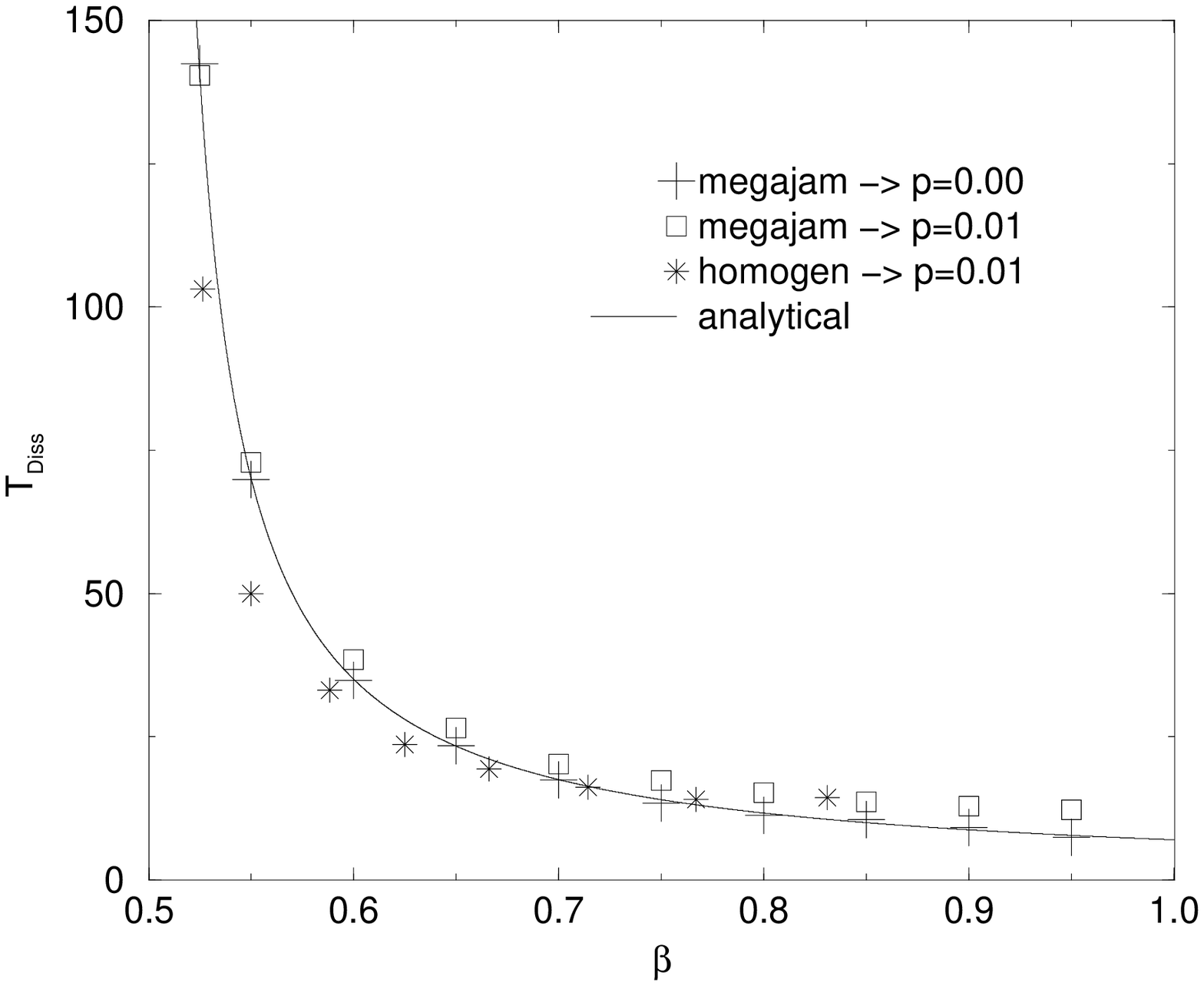,width=6.5cm}} \caption{\protect{The outflow
rate in both diagrams is set to $\alpha = 0.5$ and a width of $n_0=4$
is chosen for the induced minijam. We performed $10^{5}$ simulation
runs (induced minijams) for every data point. {\bf Left:} The
probability that the initial minijam grows to a wide jam is
illustrated for different starting conditions. The analytical curve is
given by $S=1-\Pi_{n_{0}(1)}$ (see Eqn.~(\ref{endsolution})). {\bf
Right:} The mean dissolution time of the initial minijam is shown in
the left part of the diagram.}}
\label{exp}
\end{figure}

In Fig.~\ref{exp} we show the simulated sensitivity $S = 1 -
\Pi_{n_0}(1)$ and the mean resolving time for the different initial
conditions and compare these results to the analytical predictions of
Sec.~\ref{sec_RW}. It is clear that for scenario (A) the analytical
results are exact. Even scenario (B) shows an excellent agreement with
the analytical curve, but small deviations occur due to the fact that
the inflow is no i.i.d.\ random variable anymore although the mean
inflow is still identical to that of scenario (A). Furthermore it
should be noted that for the considered dissolution of a small {\it
damage} the time scale is small, so that local deviations in the gap
distribution can play an important role. In the case of the
homogeneous initialization (C) we found larger deviations from the
analytical curve. The origin of this discrepancy is also the gap
distribution which in contrast to the megajam initializations is not
generated through a stochastic outflow parameter. Instead it is
determined by vehicle interactions due to the model dynamics
(simulation runs until relaxation). Hereby repulsive forces between
the cars lower the probability of finding large gaps. Therefore the
theory overestimates the dissolving probability in the case of a
periodic system with homogeneous starting conditions (C). Note, that
for large $p$ spontaneous jams can appear in the free flow region
before a car is able to enter the induced minijam (starting condition
(B)) or in the case of homogeneous initialization (C) jams can appear
in the system before the steady state is reached. Therefore we do not
consider initializations with higher $p$ in this work. Our aim was to
analyze the dynamics of a single jam. Thus we chose the VDR model
which exhibits phase separation but only for $p_{0}\gg p$. Nonetheless
we want to point out that the random walk approach for the dynamics of
a single jam seems to be generic for various stochastic CA models for
traffic flow.

\section{Summary}

We have analyzed the VDR model which exhibits phase separation and
meta\-stable states in order to obtain a deeper insight into the
jamming behavior in cellular automata models for traffic flow. An
analytical approach in terms of random walk theory has been suggested
to determine characteristic quantities of wide jams, especially
resolving probabilities and lifetimes. The analytical results reveal
interesting peculiarities of the model. One finds, e.g., a shift in
the convergence of the dissolving probability going from $n_0=1$ to
greater values of $n_0$. The analytical predictions are compared to
simulation results. For this purpose a damage scenario is considered
by initializing a finite single jam into an undisturbed system.
Different boundary conditions are assumed in our investigations. In
one case we choose open boundary conditions whereby the free-flow car
distribution is generated through the outflow of a sufficiently large
megajam at the left boundary. Additionally, a periodic system with
homogeneous initialization is considered. In this case the gap
distribution is generated through the dynamics of the model
(simulation are performed until the system has relaxed). For the
megajam initialization we found a very good or even exact agreement
between simulation and theory. The homogeneous initialization with
periodic boundary condition shows larger deviations from the predicted
curve but the overall agreement also rather good. We stress that the
random walk approach renders the jamming dynamics of the model. Nagel
and Paczuski \cite{nageljam} also analysed the lifetime of jams in
another stochastic CA model for traffic flow, the cruise control limit
of the NaSch model, and found good agreement with random walk
theory. Therefore we assume that the jamming behavior is generic for a
lot of the stochastic CA models for traffic flow.

Furthermore, the fact that one can interpret the emergence of wide
jams in terms of probability theory points out a main difference
between stochastic CA models for traffic flow and hydrodynamical
approaches. To explain the major differences we want to focus on the
hydrodynamical model introduced by Kerner and Konh\"auser
\cite{kerner}. Although the VDR model invokes most of the
characteristic properties (i.e., linear growth of emerging jams,
moving transition layer, ...) of jams in fluid-dynamical models the
formation of large jams due to local perturbations is completely
different. In the hydrodynamical model a wide jam is formed from an
initially homogeneous state through an external damage which exceeds a
critical size. This critical size strongly depends on the density
whereby damages below this size dissolve and damages above the
critical size lead to wide jams. In contrast, in the VDR model an
induced damage leads to a wide jam only with a certain probability.
There is no quantity like a critical value for the formation of jams
but the probability that a damage causes a jam of course depends on
the size of the damage and the density of the initial homogeneous
state. Additionally one finds spontaneous jam formation without
external influences due to velocity fluctuations in the metastable
branch of the model.

To conclude, the results presented here are of practical relevance for
various applications of traffic flow using stochastic CA models.  Due
to its simplicity this class of models has become popular for large
scale computer simulation (city or highway networks). Complex
networks usually contain many bottlenecks such as crossings, lane
reductions, traffic lights or traffic signs. Therefore induced jams
often play an important role in realistic traffic scenarios and a
proper understanding of the jamming process and dynamics is
benefitable.
 


\end{document}